\documentclass[preprint2]{aastex6}
\renewcommand{\thefootnote}{\fnsymbol{footnote}}

\newcommand{\EP}{EPIC201702477}
\newcommand{\EPb}{EPIC201702477b}
\newcommand{\kepler}{\textit{Kepler}}
\newcommand{\KK}{\textit{K2}}

\newcommand{\kms}{\ensuremath{\rm km\,s^{-1}}}
\newcommand{\ar}{\ensuremath{a/R_{\star}}}
\newcommand{\rprs}{\ensuremath{R_{BD}/R_{\star}}}
\newcommand{\Rp}{\ensuremath{R_{\rm BD}}}
\newcommand{\Mp}{\ensuremath{M_{\rm BD}}}
\newcommand{\rjup}{\ensuremath{R_{\rm J}}}
\newcommand{\mjup}{\ensuremath{M_{\rm J}}}
\newcommand{\rsun}{\ensuremath{R_{\rm \sun}}}
\newcommand{\msun}{\ensuremath{M_{\rm \sun}}}
\newcommand{\rhosun}{\ensuremath{\rho_{\rm \sun}}}
\newcommand{\masy}{\ensuremath{\rm mas\,yr^{-1}}}
\newcommand{\gcmc}{\ensuremath{\rm g\,cm^{-3}}}
\newcommand{\teff}{\ensuremath{T_{\rm eff}}}
\newcommand{\logg}{\ensuremath{\log{g}}}
\newcommand{\vsini}{\ensuremath{v \sin{i}}}
\newcommand{\feh}{\ensuremath{\rm [Fe/H]}}

\newcommand{\mplanet}{\ensuremath{66.9\pm1.7\,\mjup}}
\newcommand{\rplanet}{\ensuremath{0.757\pm0.065\,\rjup}}
\newcommand{\dplanet}{\ensuremath{191\pm51\,\gcmc}}
\newcommand{\pplanet}{\ensuremath{40.73691\pm0.00037\,\rm day}}
\newcommand{\eplanet}{\ensuremath{0.2281\pm0.0026}}

\newcommand{\vespa}{\textsc{\lowercase{vespa}}}
\newcommand{\specmatch}{\textsc{\lowercase{specmatch}}}
\newcommand{\pastis}{\textsc{\lowercase{pastis}}}
\newcommand{\exofast}{\textsc{\lowercase{exofast}}}
\newcommand{\COND}{\textsc{\lowercase{COND03}}}

\shortauthors{Bayliss et al.}
\shorttitle{
\EP
}

\begin{document}
\title{\EPb: A long period transiting Brown Dwarf from \KK}

\author{
  D.~Bayliss\altaffilmark{1,2},
  S.~Hojjatpanah\altaffilmark{3,4},
  A. Santerne\altaffilmark{3,5},
  D.~Dragomir\altaffilmark{6},
  G.~Zhou\altaffilmark{7},
  A.~Shporer\altaffilmark{8,9},
  K.~D. Col\'on\altaffilmark{10,11},
  J.~Almenara\altaffilmark{12},
  D.~J.~Armstrong\altaffilmark{13,14},
  D.~Barrado\altaffilmark{15},
  S.~C.~C.~Barros\altaffilmark{3},
  J.~Bento\altaffilmark{2},
  I.~Boisse\altaffilmark{5},
  F.~Bouchy\altaffilmark{1},
  D.~J.~A.~Brown\altaffilmark{13},
  T.~Brown\altaffilmark{16,17},
  A.~Cameron\altaffilmark{18},
  W.~D.~Cochran\altaffilmark{19},
  O.~Demangeon\altaffilmark{5},
  M.~Deleuil\altaffilmark{5},
  R.~F.~D\'iaz\altaffilmark{1},
  B.~Fulton\altaffilmark{20,21},
  K.~Horne\altaffilmark{18},
  G.~H\'ebrard\altaffilmark{22,23},
  J.~Lillo-Box\altaffilmark{24},
  C.~Lovis\altaffilmark{1},
  D.~Mawet\altaffilmark{25},
  H.~Ngo\altaffilmark{25},
  H.~Osborn\altaffilmark{13},
  E.~Palle\altaffilmark{26},
  E.~Petigura\altaffilmark{25,27},
  D.~Pollacco\altaffilmark{13},
  N.~Santos\altaffilmark{3,28},
  R.~Sefako\altaffilmark{29},
  R.~Siverd\altaffilmark{16},
  S.~G.~Sousa\altaffilmark{3},
  M.~Tsantaki\altaffilmark{3,30}
}

\altaffiltext{1}{Observatoire Astronomique de l'Universit\'e de
  Gen\`eve, 51 ch. des Maillettes, 1290 Versoix, Switzerland; email:
  daniel.bayliss@unige.ch}
\altaffiltext{2}{Research School of
  Astronomy and Astrophysics, Australian National University,
  Canberra, ACT 2611, Australia}
\altaffiltext{3}{Instituto de Astrof\'isica e Ci\^{e}ncias do Espa\c co, Universidade do Porto, CAUP, Rua das Estrelas, 4150-762 Porto, Portugal}
\altaffiltext{4}{Department of Physics, University of Zanjan, University Blvd, 45371-38791, Zanjan, Iran}
\altaffiltext{5}{Aix Marseille Universite, CNRS, Laboratoire d'Astrophysique de Marseille UMR 7326, 13388, Marseille, France}
\altaffiltext{6}{The Department of Astronomy and Astrophysics, University of Chicago, 5640 S Ellis Ave, Chicago, IL 60637, USA}
\altaffiltext{7}{Harvard-Smithsonian Center for
  Astrophysics, Cambridge, MA 02138, USA}
\altaffiltext{8}{Jet Propulsion Laboratory, California Institute of Technology, 4800 Oak Grove Drive, Pasadena, CA 91109, USA}
\altaffiltext{9}{Sagan Fellow}
\altaffiltext{10}{NASA Ames Research Center, M/S 244-30, Moffett Field, CA 94035, USA}
\altaffiltext{11}{Bay Area Environmental Research Institute, 625 2nd St. Ste 209 Petaluma, CA 94952, USA}
\altaffiltext{12}{Univ. Grenoble Alpes, IPAG, F-38000 Grenoble, France}
\altaffiltext{13}{Department of Physics, University of Warwick, Gibbet Hill Road, Coventry, CV4 7AL, UK}
\altaffiltext{14}{ARC, School of Mathematics and Physics, Queens University Belfast, University Road, Belfast BT7 1NN, UK}
\altaffiltext{15}{Depto. de Astrof\'isica, Centro de Astrobiolog\'ia (CSIC-INTA), ESAC campus, 28691 Villanueva de la Ca\~nada, Spain}
\altaffiltext{16}{Las Cumbres Observatory Global Telescope, Goleta, CA 93117, USA}
\altaffiltext{17}{CASA, Department of Astrophysical and Planetary Sciences, University of Colorado, 389-UCB, Boulder, CO 80309, USA}
\altaffiltext{18}{SUPA School of Physics and Astronomy, University of St. Andrews, KY16 9SS, UK}
\altaffiltext{19}{McDonald Observatory and Department of Astronomy, Univeristy of Texas at Austin, USA}
\altaffiltext{20}{Institute for Astronomy, University of Hawaii, 2680 Woodlawn Drive, Honolulu, HI 96822-1839, USA}
\altaffiltext{21}{NSF Graduate Research Fellow}
\altaffiltext{22}{Institut d'Astrophysique de Paris, UMR7095 CNRS, Universit\'e Pierre \& Marie Curie, 98bis boulevard Arago, 75014 Paris, France}
\altaffiltext{23}{Observatoire de Haute-Provence, Universite d'Aix-Marseille \& CNRS, 04870 Saint Michel l'Observatoire, France}
\altaffiltext{24}{European Southern Observatory (ESO), Alonso de Cordova 3107, Vitacura, Casilla 19001, Santiago de Chile, Chile}
\altaffiltext{25}{California Institute of Technology, 1200 E. California Boulevard, Pasadena, CA 91125, USA}
\altaffiltext{26}{Instituto de Astrof\'isica de Canarias, E-38205 La Laguna, Tenerife, Spain}
\altaffiltext{27}{Hubble Fellow}
\altaffiltext{28}{Departamento de F\'isica e Astronomia, Faculdade de Ci\^encias, Universidade do Porto, Rua do Campo Alegre, 4169-007 Porto, Portugal}
\altaffiltext{29}{South African Astronomical Observatory, P.O. Box 9, Observatory 7935, South Africa}
\altaffiltext{30}{Instituto de Radioastronom\'ia y Astrof\'isica, IRyA, UNAM, Campus Morelia, A.P. 3-72, C.P. 58089 Michoac\'an, Mexico}
\altaffiltext{$\dagger$}{This paper uses observations obtained with facilities of the Las Cumbres Observatory Global Telescope.  Based in part on observations made with the ESO~3.6\,m Telescope at the ESO Observatory in La Silla.}

\begin{abstract}
We report the discovery of \EPb, a transiting brown dwarf in a long period (\pplanet) and eccentric ($e$=\eplanet) orbit. This system was
initially reported as a planetary candidate based on two transit events seen in \KK\ Campaign 1 photometry and later validated as an exoplanet.  We confirm the transit and refine the ephemeris with two subsequent ground-based
detections of the transit using the LCOGT\,1\,m telescope network. We
rule out any transit timing variations above the level of $\sim$30\,s. Using high precision radial velocity measurements from HARPS and SOPHIE we identify the transiting companion as a brown dwarf with a mass, radius, and bulk density of \mplanet, \rplanet, and
\dplanet\ respectively.  \EPb\ is the smallest radius brown dwarf yet discovered, with
a mass just below the H-burning limit.  It has the highest density of any planet, substellar mass object or main-sequence star discovered so far.  We find
evidence in the set of known transiting brown dwarfs for two
populations of objects - high mass brown dwarfs and low mass brown dwarfs.  The higher-mass population have radii in very close agreement to theoretical models, and show a lower-mass limit around 60\,\mjup.  This may be the signature of mass-dependent ejection of systems during the formation process.   
\end{abstract}


\section{Introduction}
\label{sec:introduction}
The scarcity of companions with masses between $13M_{J}$ and $80M_{J}$
around main sequence stars, the ``brown dwarf desert'', was first
identified from numerous radial velocity planet searches
\citep{2000PASP..112..137M,2000A&A...355..581H}.  Radial velocity surveys combined
with astrometric data also show the brown dwarf desert to be real \citep{2011A&A...525A..95S, 2016A&A...588A.144W}.  Ground-based transit surveys, primarily
sensitive to exoplanets with radii  similar to or larger than Jupiter, seemed to
confirm this desert by finding many Jupiter-mass objects, but very few
brown dwarfs - see discoveries of WASP \citep{2006PASP..118.1407P},
HATNet \citep{2004PASP..116..266B}, HATSouth
\citep{2013PASP..125..154B}, and KELT \citep{2012PASP..124..230P}.  In fact, of this 179 transiting planets discovered by these groups, only two, WASP-30b
\citep{2011ApJ...726L..19A} and KELT-1b \citep{2012ApJ...761..123S},
have masses above 13\,\mjup.  This is despite brown dwarfs having
similar radii to hot Jupiters ($\sim$1\,\rjup) and high mass objects
being much easier to characterize with the routine radial velocity
follow-up used by these projects.  The space-based CoRoT mission \citep{1999PCEC...24..567R} discovered three transiting brown dwarfs: CoRoT-3b \citep{2008AA...491..889D}, CoRoT-15b \citep{2011AA...525A..68B} and CoRoT-33b \citep{2015AA...584A..13C}.  The Kepler mission uncovered another four transiting brown dwarfs: Kepler-39b \citep{2011AA...533A..83B}, KOI-205b \citep{2013AA...551L...9D},  KOI-415b \citep{2013AA...558L...6M}, and  KOI-189b \citep{2014AA...572A.109D}.  Additionally KOI-554b and KOI-3728b have masses, measured via light curve modulations, just above 80\,\mjup, putting them very close to the brown dwarf regime \citep{2016arXiv160602398L}.  However the bulk of planet candidates discovered by the Kepler space mission \citep{2010Sci...327..977B} have measured radii but not masses, so are not able to provide a constraint on the brown
dwarf population due to the radius degeneracy between gas giants and
brown dwarfs.  The recent radial velocity study of
\cite{2016A&A...587A..64S} was able to measure the masses for a sample
of large-radius Kepler candidates and found the occurrence rate of
brown dwarfs with periods less than 400\,days to be $0.29\pm0.17\%$.

Brown dwarfs are thought to form via gravitational instability or
molecular cloud fragmentation, whereas giant gas planets form via core
accretion \citep{2014prpl.conf..619C}. However, it is possible that
core accretion may produce super-massive planets in the 20-40\,\mjup\
range \citep{2009A&A...501.1139M}, and gravitational instability may
also form gas giant planets \citep{2015MNRAS.452.1654N}. Thus the line between gas giants and
brown dwarfs is a blurred one. It is argued that the distinction between
these objects should be linked with formation mechanisms
\citep{2014prpl.conf..619C}, and these different formation scenarios
are almost certainly responsible for the brown dwarf desert rather
than some observational bias \citep{2014MNRAS.439.2781M}.

In this paper we report the discovery of a new transiting brown dwarf,
\EPb\ ($V$=14.57), for which we can measure a precise mass and radius.  In
Section~\ref{sec:observations} we outline the photometric data from
the Kepler space telescope and the LCOGT\,1\,m network. We also describe
the spectroscopic observations used to measure the radial velocities
of \EP\ and to spectroscopically characterize the host star.  We describe the high angular resolution imaging we carried out to further rule out blend scenarios.  In
Section~\ref{sec:analysis} we carry out a joint analysis of the
observational data in order to determine the physical and orbital
characteristics of the transiting body.  Finally, in Section
\ref{sec:discussion} we look at the implications of this discovery
in terms of the known population of well characterized brown dwarfs, the mass-radius-age relationship for brown dwarfs, and the evidence for a lower mass edge to the population of high mass brown dwarfs.

\section{Observations}
\label{sec:observations}

\subsection{\KK}
\label{sec:kepler}
The NASA \kepler\ telescope is a 0.95\,m space-based Schmidt
telescope with a 105\, deg$^{2}$ field-of-view
\citep{2010Sci...327..977B}. The original mission monitored a single
field in the northern hemisphere, and was designed to determine the
frequency of Earth-like planets in the galaxy. After four years of
operations two of the spacecraft's reaction wheels failed, ending the
original mission.  However, the telescope was re-purposed to monitor
selected ecliptic fields, which optimizes the pointing stability, in a
new mission called \KK \citep{2014PASP..126..398H}.

\KK\ monitors pre-selected target stars in ecliptic fields
for durations of approximately 80 days.  While this duration is much
shorter than the original \kepler\ mission, it is still a significant
improvement over ground-based monitoring which must contend with
interruptions from poor-weather and the Earth's day-night cycle.  The
result of this is that \KK\ is currently the premier facility for
finding long period transiting planets, and \EPb\ is an example of
such a discovery.

\EP\ was monitored by \KK\ as part of Campaign 1 between 2014 May 30
and 2014 August 21.  The star was included as part of program GO1059
(Galactic Archaeology), which aimed to monitor red giant stars and
selected targets based purely on a 2MASS magnitude and color cut.  The
2MASS color of \EP\ is $J-K=0.502$, right at the edge of the color cut
for the program ($J-K>0.5$).  Given this and the magnitude of the
target ($V$=14.57), it was not likely \EP\ would be a giant star, and indeed our
spectroscopy shows the star is a Sun-like dwarf (see
Section~\ref{sec:spectype}).

\EPb\ was first identified as a transiting exoplanet candidate in
\cite{2015ApJ...806..215F}, where a transit signal with a 40.7365\,day
periodicity was reported.  The candidate was studied further by
\cite{2015ApJ...809...25M} using existing SDSS imaging, and they
noted the presence of a neighbor at 12.11$\arcsec$ with a $\Delta r
= 4.65 \pm 0.09$ mag. They concluded this neighbor was not sufficiently
close to be responsible for the transit signal identified using a
photometric aperture with a size of 10$\arcsec$.  They also calculated the
false positive probability (FPP) for \EPb\ using the \vespa\ algorithm
\citep{2012ApJ...761....6M} to be 4$\times$10$^{-3}$, and therefore deemed it
to be a ``validated planet'' (defined as FFP$<0.01$).

Due to its long orbital period there are only two transit events in
the \KK\ data, and at the \KK\ 30-minute cadence this equated to just sixteen
in-transit data points.  Such poor sampling of the transit event, even
given the exquisite precision of \KK, meant that the transit
parameters were rather poorly defined.  In such circumstances, further
ground-based photometry is very important in order to help fully
characterize the system.

Of the 37 candidates presented by \citet{2015ApJ...806..215F},
\EP\ has the longest orbital period, with the exception of
EPIC201257461, which has been shown to be a false candidate
\citep{2015ApJ...809...25M}.  The reported planet/star radius ratio of
\EPb\ is $R_{P}/R_{star}=0.0808$, indicating a gas giant exoplanet
assuming a solar-type host.

We downloaded the \KK\ pixel data for \EP\ from the Mikulski Archive for Space Telescopes (MAST)\footnote[1]{archive.stsci.edu/k2/} and used a modified version of the CoRoT imagette pipeline \citep{2015MNRAS.454.4267B} to extract the light curve. We computed an optimal aperture based on signal-to-noise of each pixel. The background was estimated using the $3\sigma$ clipped median of all the pixels in the image outside the optimal aperture and removed before aperture photometry was performed. We also calculated the centroid using the Modified Moment Method by \citet{1989AJ.....97.1227S}. For \EP\ we found that a 14 pixel photometric aperture resulted in the best photometric precision.

\begin{figure*}[tbp]
\plotone{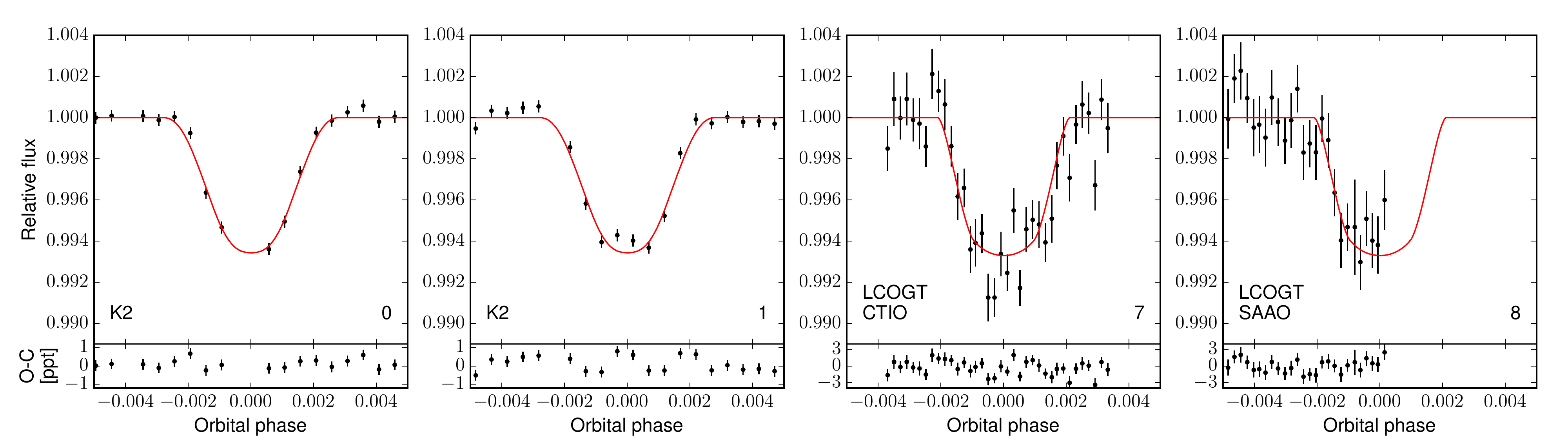}
\caption{Transit light curves for \EP\ phase-folded to the best
  fitting period of $P=\pplanet$.  Black
  circles are the photometric data-points, while the red line is the
  best-fit transit model.  First two light curves are the \KK\ data,
  comprising of two transit events in the Kepler bandpass.  Third light curve is the LCOGT\,1\,m+Sinistro $r$-band light curve from a
  single transit event observed from CTIO, Chile on 2015~March~15.
  Fourth light curve is the LCOGT\,1\,m+SBIG $r$-band light curve from a
  single transit event observed from SAAO, South Africa on 2015~April~28.
\label{fig:phot}}
\end{figure*}

The degraded pointing stability of the \KK\ mission results in flux variations correlated with the star's position on the CCD.  To correct for this we used a self-flat-fielding procedure similar to \citet{2014PASP..126..948V} that assumes the movement of the satellite is mainly in one direction.  A full description of the pipeline given in S.~Barros et al. (2015, submitted). The final light curve of \EP\ has mean out-of-transit RMS of $293\,$ ppm and the two transit events in the light curve are plotted in Fig.~\ref{fig:phot}.


\subsection{LCOGT}
\label{sec:lcogt}
The Las Cumbres Observatory Global Telescope (LCOGT) is a network of fully
automated telescopes \citep{2013PASP..125.1031B}.  Currently there are
ten LCOGT 1\,m telescopes operating as part of this network, eight of
which are in the southern hemisphere: three at the Cerro Tololo
Inter-American Observatory (CTIO) in Chile, three at the South African
Astronomical Observatory (SAAO) in South Africa, and two at Siding
Spring Observatory (SSO) in Australia.  Each telescope is equipped with
an imaging camera; either a ``Sinistro'' or an SBIG STX-16803.  The
Sinistro is LCOGT's custom built imaging camera that features a
back-illuminated 4K$\times$4K Fairchild Imaging CCD with $15\,\mu$m pixels (CCD486 BI).  With a plate scale of $0.387"/$pixel, the
Sinistro cameras deliver a FOV of $26.6~\arcmin\times26.6~\arcmin$,
which is important for monitoring a sufficient number of reference
stars for high-precision differential photometry.  The cameras are read out by four amplifiers with $1\times1$ binning, with a readout time of $\approx 45$\,s.  The SBIG
STX-16803 cameras are commercial CCD cameras which feature a
frontside-illuminated 4K$\times$4K CCD with $9\,\mu$m pixels - giving
a field of view of $15.8\arcmin\times15.8\arcmin$.  These cameras are typically read out in $2\times2$ binning mode, which results in a read-out
time of 12\,s. 

The Transiting Exoplanet CHaracterisation (TECH)\footnote[2]{lcogt.net/science/exoplanets/tech-project/} project uses the 1\,m
telescopes in the LCOGT network to photometrically characterize
transiting planets and transiting planet candidates.  A major focus of
the TECH project is to characterize long period ($>10$\,day) transiting planets or candidates which are difficult to monitor with
single site or non-automated telescope systems.  As such, \EP\ was
selected as a good candidate for photometric monitoring, and was
entered to the automated observing schedule in 2015 February.

The first transit event for \EPb\ monitored by the TECH project was on
2015 March 15 from CTIO.  We observed the target from
01:00 UT to 08:13 UT using a Sinistro in the $r$-band.  The exposure
times were 240\,s, the observing conditions were photometric, and the
airmass ranged from 2.3 to 1.2.  We detected a full transit of \EPb|
with a depth and duration consistent with that seen in the \KK\ data.
The next transit event occurred 40 days later on 2015 April 28, and was
observable from SAAO\@. \EP\ was monitored between 17:00 UT to 22:50 UT
using an SBIG camera, again in the $r$-band. The exposure times were
180\,s, the observing conditions were again photometric, and the
airmass ranged from 1.8 to 1.2.  These data show the first half
of a transit event consistent with the previous events.  The images for
both observations were calibrated via the LCOGT pipeline \citep{2013PASP..125.1031B} and aperture-photometry extracted in the standard manner as set out in \cite{2013AJ....145....5P}.  The
photometric data are provided in Table~\ref{tab:photdata}, and the
phase-folded light curves are presented in Fig~\ref{fig:phot}.


\begin{deluxetable*}{crrc}
    \tablewidth{0pc}
    \tablecaption{$r$-band Differential photometry for \EP\ from LCOGT\,1\,m\label{tab:photdata}}
    \tablehead{
      \colhead{BJD} &
      \colhead{Rel. Flux} &
      \colhead{Rel. Flux} &
      \colhead{Site/Instrument} \\
      \colhead{\hbox{~~~~(2\,400\,000$+$)~~~~}}&
      \colhead{} &
      \colhead{Error} &
      \colhead{} } \startdata
57096.5492063002& 1.0000& 0.0018&         CTIO/Sinistro\\
57096.5525186099& 1.0047& 0.0018&         CTIO/Sinistro\\
57096.5558380098& 1.0008& 0.0018&         CTIO/Sinistro\\
57096.5591604202& 1.0025& 0.0018&         CTIO/Sinistro\\
57096.5624648202& 1.0038& 0.0017&         CTIO/Sinistro\\
57096.5657806299& 1.0019& 0.0017&         CTIO/Sinistro\\
57096.5690742298& 1.0030& 0.0017&         CTIO/Sinistro\\
57096.5723725399& 1.0023& 0.0017&         CTIO/Sinistro\\
57096.5756765502& 1.0015& 0.0017&         CTIO/Sinistro\\
$ ...         $  & $   ...     $ & $   ...     $ & ...     \\
              [-1.5ex]
\enddata
\tablecomments
{
        This table is available in a machine-readable form in the
        online journal.  A portion is shown here for guidance
        regarding the format.
}  
\end{deluxetable*}

\subsection{Spectral Typing}
\label{sec:spectype}
In order to determine the stellar parameters for \EP, on 2015 March 2
we obtained a low-resolution (R=3000) spectro-photometric observation
with the Wide Field Spectrograph (WiFeS) on the Australian National
University (ANU) 2.3\,m telescope at SSO.  The methodology for this
spectral typing is fully set out in \cite{2013AJ....146..113B}.  A spectrum of R=$\lambda/\Delta
\lambda$=3000 from 3500--6000\,\AA\ is flux calibrated according to \citet{1999PASP..111.1426B} using spectrophotometric
standard stars.  We determine stellar properties, particularly \teff and \logg, via a grid search using the synthetic templates from the MARCS model atmospheres \citep{2008A&A...486..951G}.  The results
showed the star was a Sun-like dwarf star with $\teff=5600\pm200$\,K
and $\logg=4.5\pm0.5$\,dex.  Thus the transit depth was confirmed to be
consistent with a planetary-size body.

To better determine the stellar properties we obtained a spectrum of
the star with Keck/HiReS \citep{1994SPIE.2198..362V} on 2015 June 30.  The instrument was configured to the standard setup for the California Planet Search \citep{2010ApJ...721.1467H}. We collected a single 7\,min exposure using the C2 (14x0.861”) decker for a spectral resolution of R$\sim$45000 and signal-to-noise ratio of $\sim25$ per pixel at 5500\,\AA.  We used the
software \specmatch\ \citep{2015PhDT........82P} to determine the stellar properties.  The resulting parameters are listed as initial spectroscopic information in Table~\ref{tab:stellar}.  Following the methodology described in \citet{2007ApJ...664.1190S} we used these initial spectral parameters from Keck as priors for the global fitting (see Section~\ref{sec:analysis}), determined a new \logg, and then used this as a prior for a second iteration of \specmatch.   The global fit was then run again with these updated parameters, and the final solution gave $\teff = 5517 \pm 70 K$ and $\logg = 4.466 \pm 0.058$ for \EP.  The final set of stellar parameters is listed in Table \ref{tab:fitdata}.

\begin{deluxetable*}{lcl}
\tablewidth{0pc}
\tabletypesize{\scriptsize}
\tablecaption{
    Summary of stellar properties for \EP. \label{tab:stellar}
}
\tablehead{
  \multicolumn{1}{c}{Parameter} &
  \multicolumn{1}{c}{Value} &
  \multicolumn{1}{c}{Source} \\
}
\startdata
\sidehead{Identification}
R.A. (deg.)      & 175.2407940      & K2 EPIC \\
Dec. (deg.)      & +3.6815840       & K2 EPIC \\
2MASS ID.        & 11405777+0340535 & 2MASS PSC \\
\sidehead{Photometric Information}
Kepler (mag)     & 14.430           & K2 EPIC  \\
\textit{u} (mag)     & 16.312$\pm$  0.005 & SDSS DR12 \\
\textit{g} (mag)     & 14.871$\pm$  0.003 & SDSS DR12 \\
\textit{r} (mag)     & 14.354$\pm$  0.003 & SDSS DR12 \\   
\textit{i} (mag)     & 14.189$\pm$  0.003 & SDSS DR12 \\   
\textit{z} (mag)     & 14.137$\pm$  0.004 & SDSS DR12 \\
\textit{J} (mag) & 13.268 $\pm$ 0.027  & 2MASS PSC \\
\textit{H} (mag)     & 12.881 $\pm$ 0.028  & 2MASS PSC \\
\textit{K} (mag)     & 12.766 $\pm$ 0.033  & 2MASS PSC \\
\sidehead{Space Motion}
pmR.A. (\masy)          & -10.0$\pm$ 3.6 & PPMXL \\ 
pmDec (\masy)          & -9.8 $\pm$ 3.6 & PPMXL \\
mean $\gamma_{RV}$ (\kms)  & 34.0 & HARPS \\
\sidehead{Initial Spectroscopic Information}
\teff (K) & 5492 $\pm$ 60 & Keck\\
\logg & 4.12 $\pm$ 0.07 & Keck\\
\feh & -0.20 $\pm$ 0.04 & Keck\\
\vsini (\kms) &  $<$2 & Keck\\
[-1.5ex]
\enddata 
\end{deluxetable*}

\subsection{Lucky and AO Imaging}
\label{sec:imaging}
\begin{figure}[tbp]
\plotone{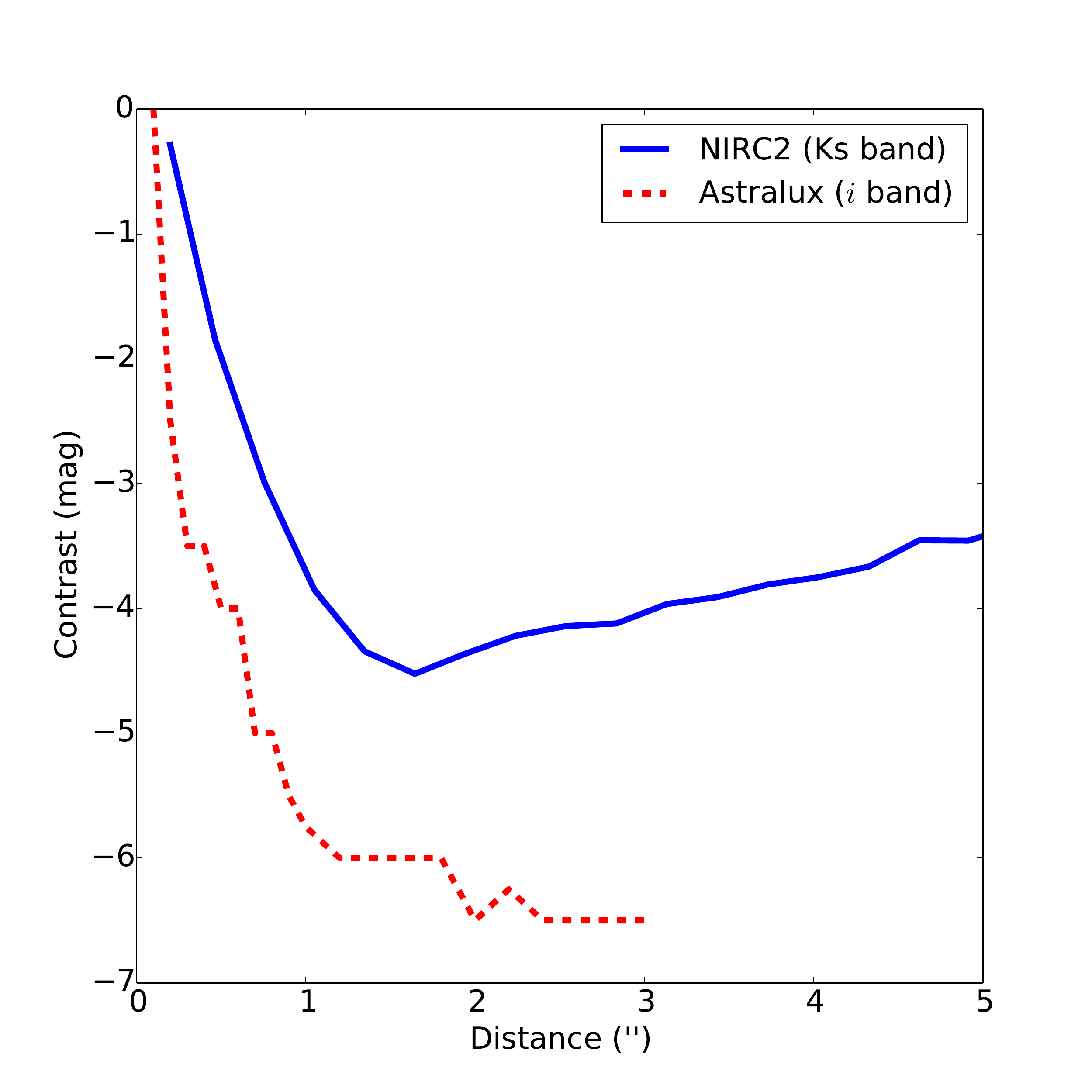}
\caption{5-sigma contrast curves for \EP\ from imaging observations.  Blue solid line:  Keck/NIRC2 K-band imaging.  Red dashed line:  Astralux lucky imaging. \label{fig:contrast}}
\end{figure}

We obtained a high-spatial resolution image with the instrument
AstraLux \citep{2008SPIE.7014E..48H}, mounted on the 2.2\,m telescope in Calar
Alto Observatory (Almer\'ia, Spain), using the lucky imaging
technique. The target was observed on  2015 November 18 under normal weather
conditions. We obtained 60000 frames with individual
exposure times of 0.060\,s, hence total exposure time of one hour, in the SDSS $i$-band. The images were reduced using the
observatory pipeline, which applies bias and flat-field correction to
the individual frames and selects the best images in terms of Strehl
ratio \citep{1902AN....158...89S}. The best 10\% of the images are then aligned
and stacked to compose the final image. The sensitivity limits are
calculated following the process explained in \citet{2014A&A...566A.103L} and
are presented in Fig.~\ref{fig:contrast}.

We observed \EP\ on 2015 December 27 using NIRC2
NGS-AO (PI: Keith Matthews) on Keck 2. We used the $Ks$ band and the narrow camera
setting.  We took a total of 4 images, each with 60 seconds of total
integration time. We calibrated the images with a flat field, dark
frames, and removed image artifacts from dead and hot pixels.  We then
created a single median-stacked image.  We do not see any stellar
companions in this image, and compute the contrast curve from the median stacked image. For every
point in the image, we compute the total flux from pixels within a box
with side length equal to the FWHM of the target star's PSF. We then
divide the image into a series of annuli with width equal to twice the
FWHM. For each annulus, we determine the $1\sigma$ contrast limit to be
the standard deviation of the total flux values for boxes inside that
annulus.  To convert from flux limits to flux ratios and differential
magnitudes, we divide the computed standard deviation by the total
flux of a similar box centered on the target
star. Figure~\ref{fig:contrast} shows the $5\sigma$ average contrast
curve.

The clear conclusion from both the lucky imaging and the AO imaging is
that the target appears to be an isolated star to within the limits
presented in our contrast curves, and this indicates the transit is occurring on the target star rather than nearby blended neighbor.

\subsection{Radial Velocities}
\label{subsec:RVs}

We performed radial velocity follow-up observations of \EP\ with the
SOPHIE \citep{2009A&A...505..853B} and HARPS \citep{2003Msngr.114...20M}
spectrographs. Both instruments are high-resolution (R $\approx$
40,000 and 110,000 for SOPHIE and HARPS, respectively), fiber-fed, and
environmentally-controlled echelle spectrographs covering
visible wavelengths.  We obtained three spectra with SOPHIE (OHP programme ID:
15B.PNP.HEBR) from 2015~June~12 to 2016~February~17 with exposure time of
1800\,s and 3600\,s, reaching a signal-to-noise ratio between 8 and 22 per pixel at 5500\,\AA. We obtained ten other spectra with
HARPS (ESO programme ID: 096.C-0657) from 2016~January~10 to 2016~February~15 with exposure time between 900\,s and 3600\,s, corresponding to
a signal-to-noise ratio between 3 and 17 per pixel at 5500\,\AA.

\begin{figure}[tbp]
  \plotone{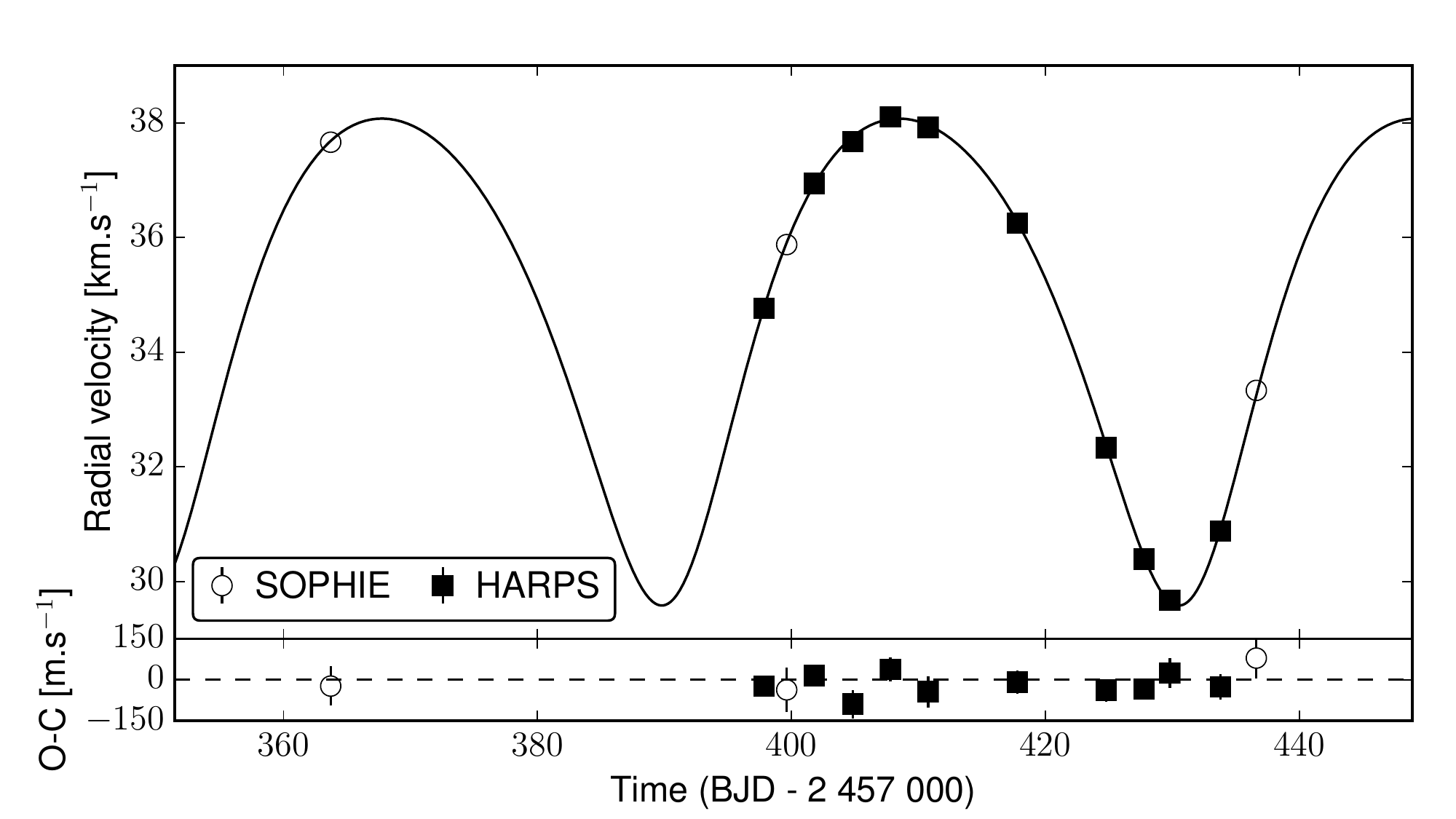}
  \plotone{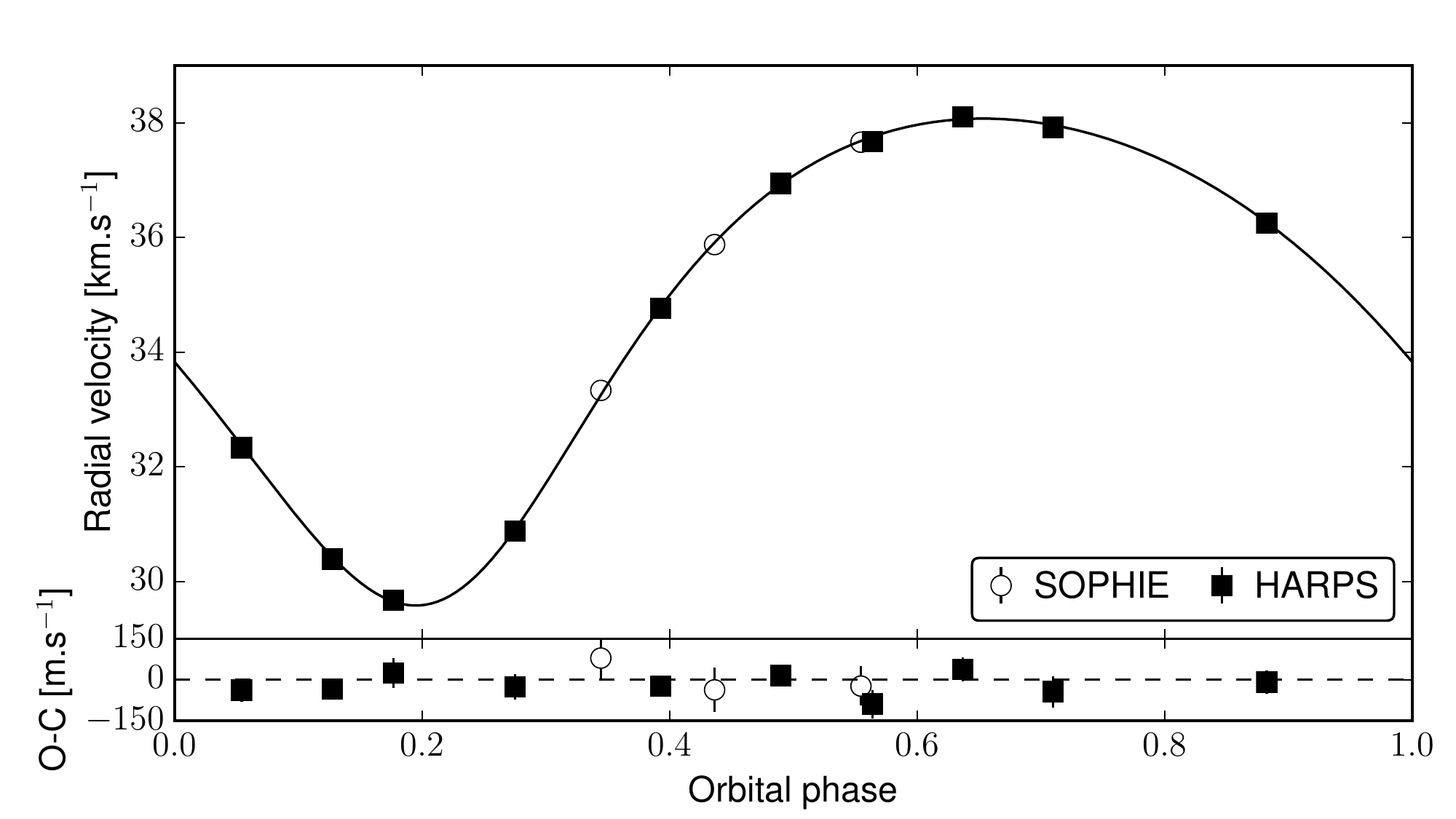}
\caption{
  \textit{Top:} Radial velocity measurements for \EP \ from the HARPS
  (solid squares) and SOPHIE (empty circles) spectrographs plotted
  against time.  The black line shows the best fit global model (see
  Section~\ref{subsec:joint}).  Lower inset panel shows O-C residuals from
  this best fit model.  \textit{Bottom:} Same as above, but
  phase-folded to the best-fit period of P=\pplanet.
\label{fig:rvs}}
\end{figure}

All spectra were reduced with the online pipeline available at the
telescopes. The spectra were then cross-correlated with a template
mask that corresponds to a G2V star \citep{1996A&AS..119..373B}. This
template was chosen to be close in spectral type to the host star.
Radial velocities, bisector span and full-width half maximum (FWHM)
were measured on the cross-correlation function and
their associated uncertainties were estimated following the methods
described in \cite{2001A&A...374..733B}, \cite{2010A&A...523A..88B},
and \cite{2015MNRAS.451.2337S}. SOPHIE radial velocities were
corrected for charge-transfer inefficiency
\citep{2009EAS....37..247B} using the equation provided in
\cite{2012A&A...545A..76S}. The derived radial velocities are
reported in Table \ref{tab:rvdata} and plotted in Fig.~\ref{fig:rvs}.

Our radial velocity measurements show a large amplitude
($K=4.252\pm0.028\,\kms$) variation in-phase with the photometric
ephemeris and indicative of a brown dwarf mass companion in an
elliptical orbit.  We use these radial velocity data to determine the
planetary parameters in Section~\ref{sec:analysis}.

\begin{deluxetable*}{cccccccccc}
\tablewidth{0pc}
\tablecaption{SOPHIE and HARPS RVs of \EP\ \label{tab:rvdata}} 
\tablehead{
\colhead{BJD} &
\colhead{RV} &
\colhead{$\sigma_{\rm RV}$} &
\colhead{V$_{\rm span}$} &
\colhead{$\sigma_{\rm V_{\rm span}}$}&
\colhead{FWHM}&
\colhead{$\sigma_{\rm FWHM}$}&
\colhead{Texp}&
\colhead{S/N}&
\colhead{Instrument}\\
\colhead{\hbox{~~~~(2\,400\,000$+$)~~~~}}&
\colhead{\kms} &
\colhead{\kms} &
\colhead{\kms} &
\colhead{\kms} &
\colhead{\kms} &
\colhead{\kms} &
\colhead{s} &
\colhead{} &
\colhead{}}
\startdata
57363.71073  &  37.566  &  0.025  &  -0.066  &  0.045  &  9.595  &  0.062  &  3600  &  21.7  & SOPHIE\\
57399.62998  &  35.780  &  0.046  &  0.103   &  0.082  &  9.614  &  0.114  &  3600  &  13.7  & SOPHIE\\
57436.62181  &  33.236  &  0.031  &  0.129   &  0.055  &  9.251  &  0.076  &  1800  &  8.2   & SOPHIE\\
57397.85193  &  34.765  &  0.011  &  -0.031  &  0.016  &  6.744  &  0.022  &  3600  &  12.0  & HARPS\\
57401.81118  &  36.943  &  0.007  &  0.002   &  0.010  &  6.709  &  0.013  &  3600  &  17.5  & HARPS\\
57404.83131  &  37.670  &  0.050  &  0.033   &  0.075  &  7.004  &  0.100  &  900   &  3.0   & HARPS\\
57407.80298  &  38.103  &  0.041  &  -0.117  &  0.061  &  6.802  &  0.082  &  1500  &  5.5   & HARPS\\
57410.77375  &  37.918  &  0.056  &  -0.091  &  0.084  &  6.311  &  0.111  &  900   &  2.9   & HARPS\\
57417.80853  &  36.254  &  0.041  &  0.108   &  0.061  &  6.574  &  0.081  &  900   &  3.9   & HARPS\\
57424.79651  &  32.335  &  0.039  &  -0.080  &  0.058  &  6.912  &  0.078  &  900   &  4.2   & HARPS\\
57427.78748  &  30.393  &  0.033  &  0.000   &  0.050  &  6.827  &  0.067  &  900   &  4.8   & HARPS\\
57429.80114  &  29.672  &  0.053  &  0.079   &  0.079  &  6.797  &  0.106  &  900   &  3.1   & HARPS\\
57433.79557  &  30.881  &  0.045  &  0.005   &  0.067  &  6.803  &  0.090  &  900   &  3.8   & HARPS\\
[-1.5ex]
\enddata
\tablecomments
{
        S/N is given per pixel at 550nm.
}
\end{deluxetable*}

\pagebreak

\section{Analysis}
\label{sec:analysis}

\subsection{Joint analysis}
\label{subsec:joint}
We analyzed the radial velocity and photometric data of \EP\ with the
Markov Chain Monte Carlo (MCMC) algorithm of the \pastis\ software,
which is fully described in \cite{2014MNRAS.441..983D}. We modelled the radial velocities with a Keplerian
orbit and the photometric data with the JKTEBOP package
\citep{2011MNRAS.417.2166S} and references therein. We chose as a
prior for the stellar parameters the values derived from the Keck
spectroscopy (Section \ref{sec:spectype}).  We used the Dartmouth
stellar evolution tracks of \cite{2008ApJS..178...89D} to derive the
stellar fundamental parameters (mass, radius, age) in the MCMC, in particular the stellar
density which was used to constrain the transit parameters given the
eccentricity constrained by the radial velocities, as in
\cite{2014A&A...571A..37S}.  We ignore pre-main sequence solutions as there is no evidence that this is a young star and the pre-main sequence stage is extremely short in duration.  We assumed uninformative priors for the
parameters, except for the orbital ephemeris for which we used the
ones provided by \cite{2015ApJ...809...25M}, the spectroscopic
parameters that we took from our spectral analysis, and
the orbital eccentricity for which we choose a Beta distribution as
recommended by \cite{2013MNRAS.434L..51K}. For the transit modelling, we used a quadratic law with coefficients taken from the interpolated table of \citet{2011A&A...529A..75C} for both the Kelpler and $r$ bandpasses and changed them at each step of the MCMC.

We ran 20 chains of $3\times10^{5}$ iterations each, with starting points randomly drawn from the
joint prior. We rejected non-converged chains based on
Kolmogorov-Smirnov test \citep{2014MNRAS.441..983D}. We then removed the burn-in of each chain
before thinning and merging them. We ended with more than 3000
independent samples of the posterior distribution that we used to
derive the value and 68.3\% uncertainty of each parameters that we
report in Table \ref{tab:fitdata}.

We also modelled the system independently (but with the
same datasets) using the \exofast\ software
\citep{2013PASP..125...83E}.  We find parameters and uncertainties in
close agreement with those that were derived using {\pastis}, and
therefore we only report the {\pastis} results.

\begin{deluxetable*}{lc}
\tablewidth{0pc}
\tablecaption{Parameters from Global Fit for \EP\ system\label{tab:fitdata}} 
\tablehead{
\colhead{Parameter} &
\colhead{Value}
}
\startdata
\sidehead{~~~~~~~~~~Brown Dwarf}
$P$ (days) & 40.73691 $\pm$ 0.00037\\
$T_{0}$ (BJD) & 2456811.5462 $\pm$ 0.0011\\
$T_{14}$ (hours) & 4.04 $\pm$ 0.13\\
\ar & 54.0 $\pm$ 3.4\\
\rprs & 0.0862 $\pm$ 0.0024\\
$b$ & 0.851 $\pm$ 0.023\\
$b_{sec}$ & 0.752 $\pm$ 0.023\\
$i$ (degrees) & 89.105 $\pm$ 0.082\\
$e$ & 0.2281 $\pm$ 0.0026\\
$\omega$ (degrees) & 195.9 $\pm$ 1.8 \\
$\gamma_{RV}$ (\kms) & 34.745 $\pm$ 0.020\\
$K$ (\kms)  & 4.252 $\pm$ 0.028 \\
\Mp (\mjup) & 66.9 $\pm$ 1.7 \\
\Rp (\rjup) & 0.757 $\pm$ 0.065 \\
$a$ (AU) & 0.2265 $\pm$ 0.0026\\
$\rho_{c}$ (\gcmc) & 191 $\pm$ 51 \\
\sidehead{~~~~~~~~~~Star}
\logg & 4.466 $\pm$ 0.058\\
\teff (K) & 5517 $\pm$ 70\\
\feh & -0.164 $\pm$ 0.053\\
$R_{\star}$ (\rsun) & 0.901 $\pm$ 0.057 \\
$M_{\star}$ (\msun) & 0.870 $\pm$ 0.031 \\
$\rho_{\star}$ (\rhosun) & 1.18 $\pm$ 0.24\\
age (Gyr)& 8.8 $\pm$ 4.1\\
\sidehead{~~~~~~~~~~RV and Photometry}
HARPS jitter (\kms) & $0.035^{+0.031}_{-0.018}$ \\
SOPHIE jitter (\kms) & $0.101^{+0.180}_{-0.070}$ \\
SOPHIE offset relative to HARPS (\kms) & 0.078 $\pm$ 0.081\\
K2 contamination & $0.0071^{+0.0072}_{-0.0049}$\\
K2 flux out of transit & 1.000022 $\pm$ 3.4e-05\\
K2 jitter & 0.000253 $\pm$ 2.8e-05\\
SAAO contamination & $0.030^{+0.030}_{-0.021}$\\
SAAO flux out of transit & 0.99975 $\pm$ 2.7e-04\\
SAAO jitter & 0.00039 $\pm$ 3.8e-04\\
CTIO contamination & $0.025^{+ 0.028}_{- 0.018}$\\
CTIO flux out of transit & 0.99966 $\pm$ 2.0e-04\\
CTIO jitter & 0.00089 $\pm$ 3.2e-04\\
\enddata
\end{deluxetable*}

\subsection{TTV analysis}
\label{subsec:TTV}

\begin{figure}[tbp]
\plotone{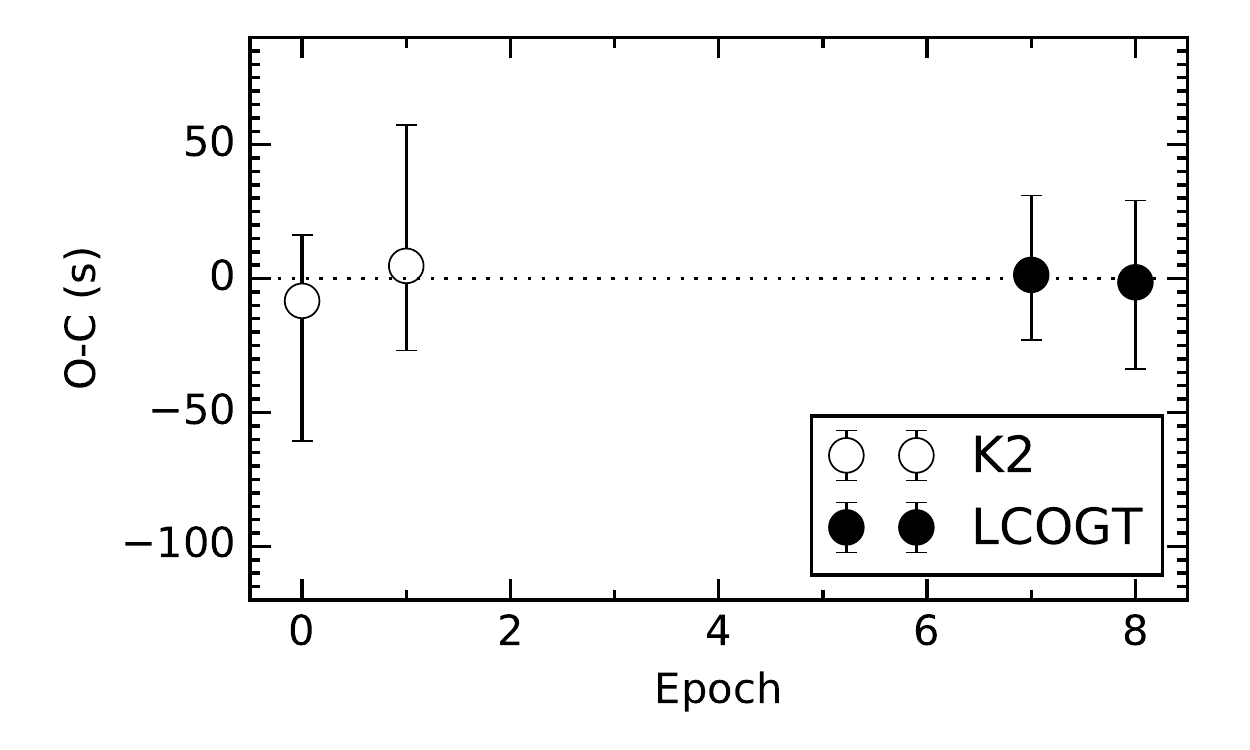}
\caption{
  Transit timing variations for \EPb\ for four transits (epochs 0 and 1
  from \KK\ data, epochs 7 and 8 from LCOGT data).  The dotted line indicates
  the mean $O-C$ offset.  We do not observe any variation at the level
  of $\sim30$\,s. 
\label{fig:ttv}}
\end{figure}

In order to test for transit timing variations (TTVs), we perform an independent fit of the K2 and LCOGT transit light curves. We fit for independent centroids $T_0$ for each transit, while forcing the transits to share the geometric parameters \ar, \rprs, and $i$. Since ground-based photometry suffers from instrumental systematics that can bias the centroid measurements, we simultaneously detrend the LCOGT light curves against a linear combination of the terms describing the time, $X$, $Y$ pixel drift, airmass trend, sky background flux, and target star FWHM variations. No significant TTVs were detected at the 30\,s level. The high cadence LCOGT light curves offer similar timing precisions as the long cadence K2 observations, and demonstrate the power of follow-up observations for long period candidates from K2. The variations in the transit centroid times are shown in Figure~\ref{fig:ttv} and listed in Table~\ref{tab:photometry}.


\begin{deluxetable*}{lccc}
\tablewidth{0pc}
\tabletypesize{\scriptsize}
\tablecaption{
    Summary of photometric observations for \EP. \label{tab:photometry}
}
\tablehead{
  \colhead{Instrument} &
  \colhead{Epoch} &
  \colhead{Transit centroid (BJD-TDB)} &
  \colhead{Filter} \\
}
\startdata
Kepler      &0& $2456811.54499\,\left(_{-60}^{+28} \right)$    & Kep. \\
Kepler      &1& $2456852.28205\,\left(_{-37}^{+61} \right)$       & Kep. \\
LCOGT\,1\,m+Sinistro &7&  $2457096.70347\,\left(_{-28}^{+34} \right)$      & sloan-r \\
LCOGT\,1\,m+SBIG &8   & $2457137.44035\,\left(_{-38}^{+35} \right)$    & sloan-r \\
[-1.5ex]
\enddata 
\end{deluxetable*}

\subsection{Out-of-transit light curve analysis}
\label{subsec:OOT}

We can place an upper limit on the companion's luminosity based on the secondary
eclipse measurements. We checked for the presence of a secondary eclipse in the \KK\ light curves; the phase of the eclipse is constrained by a Gaussian prior on the $e$ and $\omega$ orbital parameters, determined from the RV observations and presented in Table~\ref{tab:fitdata}. No secondary eclipse is detected at a $2\sigma$ upper limit of 1.96 mmag, equating to a maximum black-body temperature for the brown dwarf of $T_\mathrm{eff} < 3950\,$K.

\section{Discussion}
\label{sec:discussion}
With a period just over 40\,days, \EPb\ is the second longest period transiting brown dwarf discovered to date.  The discovery of long-period transiting systems from the \KK\ data is
encouraging, as such systems are extremely difficult to find from
ground-based surveys; HATS-17b \citep{2016AJ....151...89B} being the
current record at 16.3\,days.  Long-period systems will remain
difficult to discover even when the TESS mission is operating
\citep{2014SPIE.9143E..20R} as most fields in this survey will only be
monitored for 27\,days.  \EPb\ also demonstrates that like the Kepler mission, some fraction of the \KK\ validated planets may turn out not to be planets, even at radii down to 0.75\,\rjup, due to confusion with brown dwarf companions. 

\subsection{Populating the Brown Dwarf Desert}
\label{subsec:bd}
Including \EPb, there are just 12 known brown dwarfs ($13\mjup<\Mp<80\mjup$) that transit main sequence stars - see Table~\ref{tab:bd} for a list and \citet{2015AA...584A..13C} for a detailed list of these systems. These systems are extremely important as they provide an
independent check on the radial velocity statistics for brown dwarfs,
in addition to giving us true masses and radii.  While a full
statistical analysis is beyond the scope of this paper, we note that
from the \KK\ survey alone there have been five previously unknown hot Jupiter
discoveries (NASA Exoplanet Archive on 2016 April 20), but \EPb\ is
the first brown dwarf discovery.  Although this is in line with the relative statistics
for these two populations presented in \cite{2016A&A...587A..64S}, we caution that the target selection process for \KK\ imprints a strong bias on the sample and makes robust statistics dependent on careful modelling of the selection effects.  In addition, the detection of a large radial velocity variation may prompt follow-up efforts to be discontinued for some planet search programs.

\subsection{Two Populations of Brown Dwarfs}
\label{subsec:pops}
\cite{2014MNRAS.439.2781M} have suggested that there exist two
populations of brown dwarfs.  The first are brown dwarfs below
$\sim45$\,\mjup\ that are formed in the protoplanetary disc via
gravitational instability.  The second are brown dwarfs above $\sim45$\,\mjup\ that are formed through molecular cloud
fragmentation; essentially the very lowest mass objects of the
star-formation process.  This division of the brown dwarf population
at $\sim45$\,\mjup\ coincides with the minimum of the companion mass function derived by \cite{2006ApJ...640.1051G} and the void in the
mass range as derived from the CORALIE RV survey \citep{2011A&A...525A..95S}.
Under this division, \EPb\ would clearly be classed in the second
category as likely to be formed via molecular cloud fragmentation, as
at \mplanet\ its mass lies well above the mass division.

\begin{figure}[tbp]
\plotone{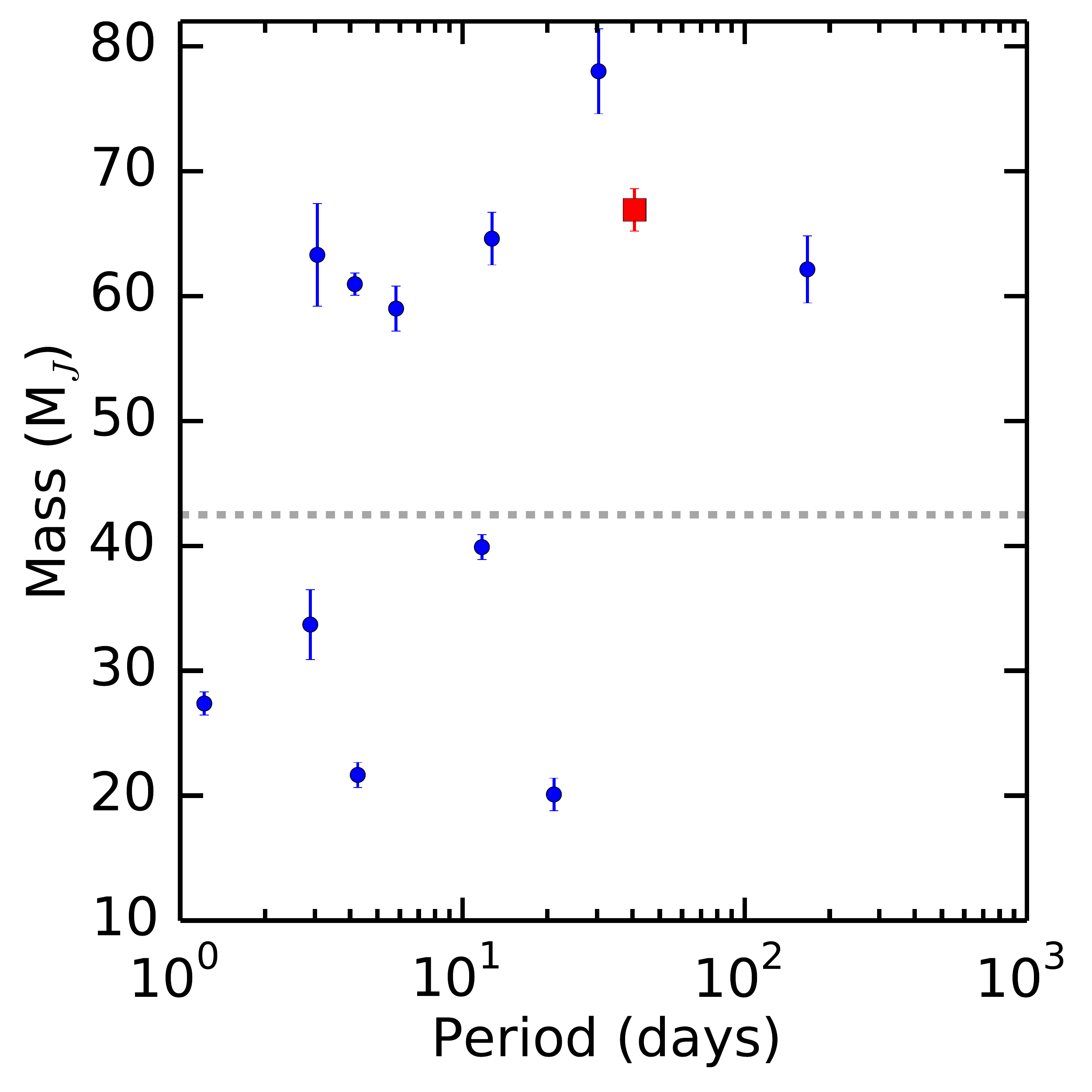}
\caption{
The masses of all known brown dwarfs that transit main sequence host
stars plotted against their orbital periods.  Blue circles are from the
literature (see Table~\ref{tab:bd}), while the red
square is \EPb. We note that \EPb\ has the second longest period of all these discoveries.  The dashed grey line indicates the 42.5\mjup\ mass at
which \cite{2014MNRAS.439.2781M} report a gap in the mass
distribution.  Based on these transiting systems alone, we do indeed
see evidence for such a gap with roughly equal numbers of companions
discovered in each population.
\label{fig:bd-mass}}
\end{figure}

Unlike pure RV detections, transiting brown dwarfs can have true
masses determined, as opposed to minimum masses.  We can also be
fairly certain that these discoveries are free from a mass bias, as
to first order the discoveries are made on the basis of the planet-to-star radius
ratio alone, and radius of the companion is largely independent of the
mass in the brown dwarf regime.  Therefore while the numbers are still
small, transiting brown dwarfs provide a critical test of the two
population model proposed in \cite{2014MNRAS.439.2781M}.  As can be
seen from Fig.~\ref{fig:bd-mass}, we do indeed see evidence of a gap
in the mass distribution between about 40\,\mjup\ and 55\,\mjup,
lending support to the two population hypothesis.

\subsection{Mass-Radius-Age Relationship for Brown Dwarfs}
\label{subsec:mrr}
\begin{figure}[tbp]
\plotone{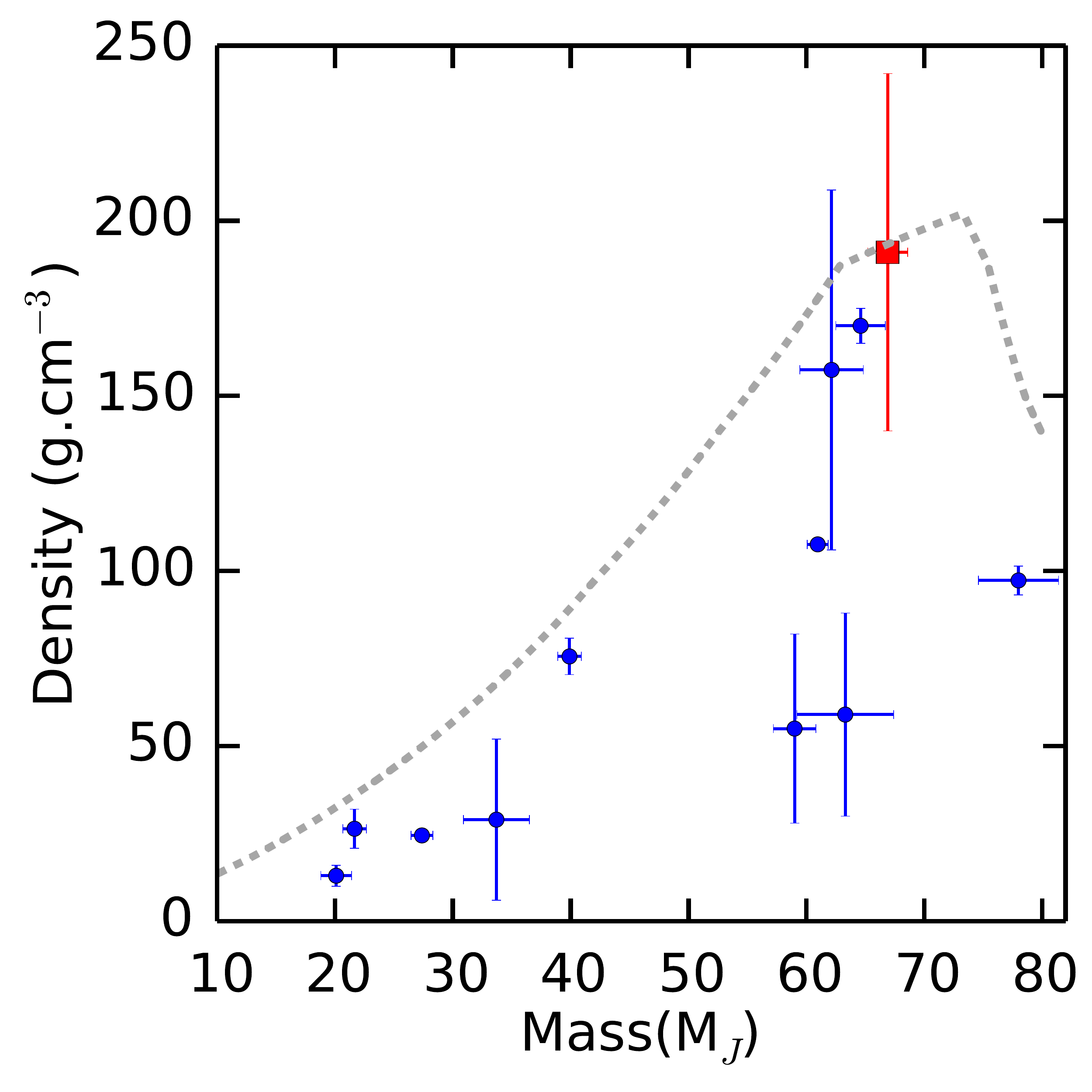}
\caption{
  The density-mass relationship for the known transiting brown dwarfs.  Sample and point symbols as for Fig.~\ref{fig:bd-mass}.  The grey dashed line indicates the \COND\ model densities for brown dwarf of 8.8\,Gyr - the estimated age of \EP.  We note that  \EPb\ stands out as the highest density object yet discovered, very near to the peak density predicted by the model.  \EPb\ has a density in perfect agreement with the 8.8\,Gyr \COND\ models. 
\label{fig:bd-density}}
\end{figure}

\begin{figure}[tbp]
\plotone{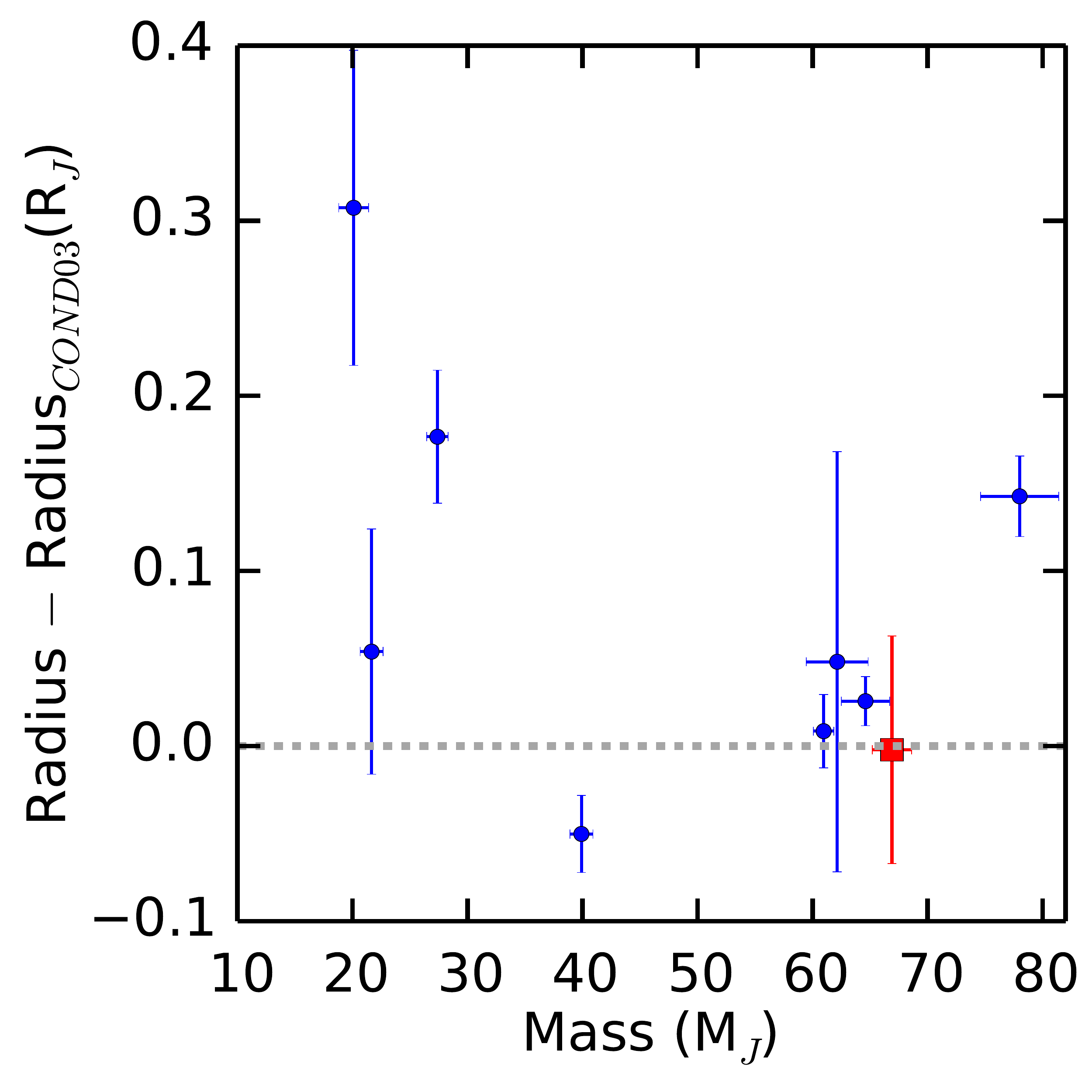}
\caption{
  The residuals between the measured brown dwarf radius and the \COND\
  model radius \citep{2003A&A...402..701B} plotted against the brown dwarf mass.  Sample and point symbols as for Fig.~\ref{fig:bd-mass}, except that we only take systems which have well determined masseses and radii (uncertainties $<$20\%).  Grey dashed-line indicates radii in perfect agreement with the \COND\ models.  We see the higher mass brown dwarfs, especially those between 60-70\,\mjup, agree very well with the \COND\ models, while lower mass systems appear to be inflated as compared to these models.  
\label{fig:bd-radius}}
\end{figure}

\EPb\ lies at the minimum for brown dwarf radii, and with a density of \dplanet\ it is the
highest density object ever discovered in the regime from planets to
main sequence stars - see Fig.~\ref{fig:bd-density}.  To investigate the mass-radius relationship for
brown dwarfs we take the known systems with precise (uncertainties $<$20\%)
mass and radius and compare the measured radius with the radius
predicted from the \COND\ models \citep{2003A&A...402..701B}.  We use the published masses and ages for each transiting brown dwarf (set out in Table~\ref{tab:bd}), and compute a \COND\ model radius for each object based on a 2-D linear interpolation of the model grid-points. We plot the difference between the measured radius and these computed radii in Fig.~\ref{fig:bd-radius}.  For hot
Jupiters, there exists a population of inflated radius objects at short
periods where the insolation flux exceeds $10^8$\,erg\,cm$^{-2}$\,s$^{-1}$ \citep{2011ApJS..197...12D}.  However for brown dwarfs the radii do not appear to exhibit such a trend, and the radii appear to be uncorrelated with the insolation flux (or for that matter orbital period).  This may be expected as most of the mechanisms proposed for giant planet inflation do not apply to these more massive brown dwarfs \citep{2011AA...525A..68B}.  A possible exception may be KELT-1b \citep{2012ApJ...761..123S} which receives extremely high insolation of 7.81$\times$10$^{9}$\,erg\,cm$^{-2}$\,s$^{-1}$ and indeed appears to be inflated. However we do note that the higher mass population of brown dwarfs are in much closer agreement to the \COND\ models than the lower mass population of brown dwarfs (see Fig.~\ref{fig:bd-radius}).

\subsection{The Mass Edge at 60\mjup}
\label{subsec:60mjup}
Of the 12 known transiting brown dwarfs, six have masses in the range of 59-67 \mjup, as shown in Fig.~\ref{fig:bd-mass}.  The lack of higher mass objects is only because we restricted our sample to objects less than 80\,\mjup (the usual limit for what is considered a brown dwarf).  Many transit and radial velocity surveys may also not report objects above this mass.  However the lack of discoveries with masses below this group of high mass transiting brown dwarfs is interesting, and appears as a sharp lower mass edge to the high-mass brown dwarfs.  While we caution that the sample size is still small, the edge is intriguing and may be related to the ejection process during formation.  In the simulations of \cite{2009MNRAS.392..413S} it is found that although the formation of brown dwarfs is approximately flat in the regime of 15-80\,\mjup, the subsequent ejection process, which results in the loss of over half of the companions, is strongly mass dependent.  Primarily it is the lower-mass brown dwarfs that are ejected, leaving behind a higher-mass population.  These simulations even show that companions around 70\,\mjup\ are among the least likely to get ejected \citep[see Fig.~15 of][]{2009MNRAS.392..413S}.  It is possible it is these objects that we find as the population of transiting brown dwarfs with masses from 60-70\,\mjup.

\begin{deluxetable*}{cccccc}
\tablewidth{0pc}
\tablecaption{Brown Dwarfs Transiting Main Sequence Stars \label{tab:bd}}
\tablehead{
  \colhead{Name} &
  \colhead{Period} &
  \colhead{Mass} &
  \colhead{Radius} &
  \colhead{Age} &
  \colhead{Ref.}\\
  \colhead{} &
  \colhead{(days)} &
  \colhead{(\mjup)} &
  \colhead{(\rjup)} &
  \colhead{(Gyr)} &
  \colhead{}\\
}
\startdata
CoRoT-3b & 4.256     & $21.66\pm1.0$    & $1.01\pm0.07$& 2.2 &{\cite{2008AA...491..889D}}\\
NLTT41135b & 2.889   & $33.7\pm2.8$     & $1.13\pm0.27$& 5.0 &{\cite{2010ApJ...718.1353I}}\\
CoRoT-15b & 3.060    & $63.3\pm4.1$     & $1.12\pm0.30$& 2.24&{\cite{2011AA...525A..68B}}\\
WASP-30b & 4.156     & $60.96\pm0.89$   & $0.889\pm0.021$& 1.5&{\cite{2011ApJ...726L..19A}}\\
LHS6343C & 12.713    & $64.6\pm2.1$     & $0.798\pm0.014$& 5.0&{\cite{2011ApJ...730...79J}}\\
Kepler-39b & 21.087  & $20.1\pm1.3$     & $1.24\pm0.10$& 4.75 &{\cite{2011AA...533A..83B}}\\
KELT-1b & 1.217     &  $27.3\pm0.93$    & $1.116\pm0.038$& 1.75&{\cite{2012ApJ...761..123S}}\\
KOI-205b & 11.720    & $39.9\pm1.0$     & $0.807\pm0.022$&  3.9&{\cite{2013AA...551L...9D}}\\
KOI-415b & 166.788   & $62.14\pm2.69$   & $0.79\pm0.12$& 10.5&{\cite{2013AA...558L...6M}}\\
KOI-189b & 30.360    & $78.0\pm3.4$     & $0.998\pm0.023$& 6.1 &{\cite{2014AA...572A.109D}}\\
CoRoT-33b & 5.819    & $59.0\pm1.8$     & $1.10\pm0.53$& 7.8&{\cite{2015AA...584A..13C}}\\
EPIC201702477b & 40.737  & $66.9\pm1.7$ & $0.757\pm0.065$& 8.8 &{this work}\\
\enddata 
\end{deluxetable*}

\clearpage
\acknowledgements

\paragraph{Acknowledgments}
This work has been carried out within the framework of the National Centre for Competence in Research "PlanetS" supported by the Swiss National Science Foundation (SNSF).
This paper includes data collected by the K2 mission. Funding for the K2 mission is provided by the NASA Science Mission directorate.
This paper makes use of data and services from NASA Exoplanet
Archive \citep{2013PASP..125..989A}, which is operated by the California
Institute of Technology, under contract with the National
Aeronautics and Space Administration under the Exoplanet Exploration
Program.
We are grateful to our colleagues who have performed some of the observations presented here with the HARPS spectrograph: F. Motalebi, A. Wyttenbach, and B. Lavie.
The Porto group acknowledges the support from the Funda\c{c}\~ao para a Ci\^encia e Tecnologia, FCT  (Portugal) in the form of the grants, projects, and contracts UID/FIS/04434/2013 (POCI-01-0145-FEDER-007672), PTDC/FIS-AST/1526/2014 (POCI-01-0145-FEDER-016886), IF/00169/2012, IF/00028/2014, IF/01312/2014 and POPH/FSE (EC) by FEDER funding through the Programa Operacional de Factores de Competitividade - COMPETE".

Partly based on observations made at Observatoire de Haute Provence (CNRS), France and with ESO Telescopes at the La Silla Paranal Observatory under programme ID 096.C-0657.
Some of the data presented herein were obtained at the W.M.~Keck Observatory, which is operated as a scientific partnership among the California Institute of Technology, the University of California and the National Aeronautics and Space Administration. The Observatory was made possible by the generous financial support of the W.M.~Keck Foundation.  The authors wish to recognize and acknowledge the very significant cultural role and reverence that the summit of Mauna Kea has always had within the indigenous Hawaiian community.

AS is supported by the European Union under a Marie Curie Intra-European Fellowship for Career Development with reference FP7-PEOPLE-2013-IEF, number 627202. J.~L-B acknowledges financial support from the Marie Curie Actions of the European Commission (FP7-COFUND) and the Spanish grant AYA2012- 38897-C02-01. JMA acknowledges funding from the European Research Council under the ERC Grant Agreement n. 337591-ExTrA. D.J.A. and D.P acknowledge funding from the European Union Seventh Framework programme (FP7/2007- 2013) under grant agreement No. 313014 (ETAEARTH). OD acknowledges support by CNES through contract 567133.
KH and ACC acknowledge support from UK Science and Technology Facilities Council (STFC) grant ST/M001296/1.
DJAB acknowledges support from the UKSA and the University of Warwick.
B.J.F. notes that this material is based upon work supported by the National Science Foundation Graduate Research Fellowship under grant No. 2014184874. Any opinion, findings, and conclusions or recommendations expressed in this material are those of the authors(s) and do not necessarily reflect the views of the National Science Foundation.
W.D.C. acknowledges support from NASA Grants NNX15AV58G and NNX16AE70G.
This material is based upon work supported by the National Science Foundation Graduate Research Fellowship under Grant No.~DGE-1144469
This work was performed in part under contract with the Jet Propulsion Laboratory (JPL) funded by NASA through the Sagan Fellowship Program executed by the NASA Exoplanet Science Institute.

\facility{CAO:2.2\,m (AstraLux), ESO:3.6\,m (HARPS), \KK, Keck:II (NIRC2), Keck:I (HIRES), LCOGT, OHP:1.93\,m (SOPHIE), ANU:2.3\,m (WiFeS)}


\end{document}